# Dynamic Motifs of Strategies in Prisoner's Dilemma Games

**Young Jin Kim, Myungkyoon Roh, Seon-Young Jeong, and Seung-Woo Son**

*Department of Applied Physics, Hanyang University, Ansan 426-791, Korea*

We investigate the win-lose relations between strategies of iterated prisoner's dilemma games by using a directed network concept to display the replicator dynamics results. In the giant strongly-connected component of the win/lose network, we find win-lose circulations similar to rock-paper-scissors and analyze the fixed point and its stability. Applying the network motif concept, we introduce dynamic motifs, which describe the population dynamics relations among the three strategies. Through exact enumeration, we find 22 dynamic motifs and display their phase portraits. Visualization using directed networks and motif analysis is a useful method to make complex dynamic behavior simple in order to understand it more intuitively. Dynamic motifs can be building blocks for dynamic behavior among strategies when they are applied to other types of games.

Email: sonswoo@hanyang.ac.kr

Fax: +82-31-400-5457

# I. INTRODUCTION

The prisoner's dilemma is well-known in game theory as a scenario that shows how two "rational" individuals may not cooperate, even though doing so would be to their benefit [1−3]. The story starts from a situation where two criminals are arrested and imprisoned. They have no way to talk to each other because they are in solitary confinement. The prosecutor offers each one a deal separately: If one confesses and the other denies the crime, the confessor will be set free, and the other will serve five years in prison. If both deny the crime and keep silent, they will each serve two years in prison. If both confess, each will serve four years in prison. The question then becomes what the choice of a rational person would be. Before we consider an individual's choice, let us think about the total punishment by summing both sentences. If the prisoners deny their guilt and keep silent, the total sentence is four years, but it is eight years if both confess, and five years if they choose different decisions. Therefore, the global optimum is for both prisoners to deny their guilt if we consider the total amount of punishment. However, from an individual viewpoint, it is always better to confess regardless of the partner's choice. Therefore, they will finally get an eight-year sentence in total even though it is the worst choice when the total amount of punishment is considered. Therefore, this is called the prisoner's dilemma. The local (individual) optimization ends up with the worst result through global (total sentence) optimization [3, 4].

However, repeating this game can have a different outcome from a single-round game because lessons toward mutual cooperation are learned. This version is called the iterated prisoner's dilemma game (IPDG) [3, 5−7]. There are three famous cooperating strategies in the IPDG: One is the "grim trigger" (GT) who cooperates at first, but betrays or defects forever after the opponent's single defection [8]. The second is the "tit-for-tat" (TFT), which literally does what the opponent did after initial cooperation [1−3]. The last is the "Pavlov." This strategy keeps the last decision if the result was good, but otherwise switches to a different decision. Therefore, this is also called the "win-stay, lose-switch" strategy [9]. These three strategies sound reasonable; moreover, they are very simple as they are based only on the previous decision, which means a single-step memory is sufficient to apply these strategies. The one-step memory size IPDG has 32 strategies, and their evolutionary dynamics are described by 32 first-order nonlinear differential equations. However, the relations among these 32 strategies are not fully understood [6, 7].

In this paper, we study the relations between the strategies of the IPDG with a one-step memory size by using a directed network concept to display the replicator dynamics results. The directed network concept is useful for providing a graphical representation of the relations among the strategies, which provides an understanding of the dynamics between the two strategies and an intuition for more complicated dynamics between all the available strategies. In Section II, we introduce the IPDG with 25 single-memory, non-duplicated strategies and replicator dynamics (RD). How the win-lose directed network is constructed is explained, and an analysis of the dynamic motifs is given. Section III shows the resulting win-lose network and the win-lose circulations. The results of RD among the three strategies are analyzed, and 22 existent dynamic motifs are displayed. In the last section, summaries and discussions are included.

## II. METHODS

In prisoner's dilemma (PD) games, if two players cooperate (denoted by "C") with each other, they get a payoff "R" as a reward. However, if one defects (denoted by "D") while the other cooperates, the defector receives a payoff "T" meaning a temptation, and the opponent gets a payoff "S," meaning a sucker's payoff. If both players defect, they receive a payoff "P" as a punishment. Each payoff must satisfy the conditions $T > R > P > S$ and $2R > T + S$ for the game to be a PD game [1−3]. We summarize the payoff matrix in Table 1.

When the game is repeated, various strategies are possible as player's reactions against the opponent's past actions if the player remembers the past. For example, if the player remembers only the previous action of the opponent, the player is able to react in 4 ($= 2^2$) ways, C or D, the two ways for the partner's C and the other two ways for D. If we consider a cold start when no record is possible as it is the first move, the number of possible reactions is 8 ($= 2^3$). In a similar way, if we also consider the previous action of the player along with the opponent's action, the total number of possible strategies is 32 ($= 2^5$) for mutual cooperation between the player and the opponent (denoted as |CC|). In the same way, |CD| represents a player cooperating but the opponent defecting. Likewise, |DC| and |DD| are denoted. If a convenient bitwise representation is used for these 32 cases, each strategy can be represented by using a five-bit binary system $a_0|a_1a_2a_3a_4$, where the first bit $a_0$ is the first move of the cold start, and the $a_1a_2a_3a_4$ represents the reactions for the states |CC|, |CD|, |DC|, and |DD|, respectively [6, 7]. The 32 possible strategies in the IPDG are summarized in Table 2. A decimal number is assigned for each strategy by interpreting the binary notation C and D as $1_{(2)}$ and $0_{(2)}$. For example, the 0th strategy, which corresponds to $00000_{(2)}$ ($a_4a_3a_2a_1a_0$), means D|DDDD, i.e., always defeat. The 11th strategy, which is called "tit for tat" (TFT), is $01011_{(2)}$ representing C|CDCD; first cooperate, then react as the opponent did. Even though possible representations can be done in 32 ways, seven strategies are effectively the same, and only 25 strategies are non-duplicated where stochastic noises, like mutation and simple mistakes, are ignored [7]. In this study, we consider only the 25 strategies. The seven other unused strategies are marked in red in Table 2.

1. **Evolutionary Replicator Dynamics**

Replicator dynamics (RD) is a widely used method in game theory to find an evolutionarily stable strategy (ESS) [3, 10, 11]. The population density of strategy $i$ is denoted by $x_i$, and its time derivative $\dot{x}_i$ is given by $\dot{x}_i = x_i[f_i(\vec{x}) - \bar{f}(\vec{x})]$, where $f_i(\vec{x})$ is the fitness of strategy $i$ for a given population distribution $\vec{x}$, and the mean fitness is $\bar{f}(\vec{x}) = \sum_{j=1}^{n} x_j f_j(\vec{x})$. The fitness of a certain strategy is calculated by using $f_i(\vec{x}) = \sum_{j=1}^{n} x_j P_{ij}$, where $P_{ij}$ is the average payoff of strategy $i$ from strategy $j$ per round when the two strategies are used in the IPDG [6, 7]. The average payoff matrix of strategy **P** is a 25 by 25 asymmetric matrix [7].

When two individuals play the IPDG by using their own strategies with a memory size of one, there are only four states |CC|, |CD|, |DC|, and |DD| [6, 7]. As an example, let us consider when a player makes use of strategy No. 11, TFT, and the opponent plays strategy No. 21, tree frog, as shown in Table 2. The two players' state starts from |CC|, i.e., both cooperate, then moves to |CD|, subsequently following their own strategies. Then, |DD| and |DC| follow, and their states return to |CC|, as shown in Fig. 1(a). They repeat this sequence in period-four like a recurrent state. For one period, the TFT player receives the payoff sum of R, S, P, and T while the opponent, tree frog, gets the sum of R, T, P, and S. As another example, when strategy No. 3, grim trigger, meets strategy No. 24, CD repeater, the state enters a period-two recurrent state, |DC|↔|DD|, and after the transient states |CC|→|CD|, as shown in Fig. 1(b). Because the possible periods are two, three, or four, the rounds of their least common multiple, 12, are sufficient to calculate the average payoff between the two strategies after several transient periods have been discarded. In this study, we use a payoff matrix with $T = 5, R = 3, P = 1,$ and $S = 0$, which is a typical one and satisfies the Pareto efficiency conditions [3, 5−7]. In our simulation, we use the first-order Euler method to solve the differential equation with a time step $dt = 0.01$, and we have checked that this time step size is precise enough for this numerical integration because the effective potential of this system is not stiff at all.

**2. RD of One-to-one Matches among the Strategies**

With the 25 non-duplicated strategies, we enumerate all possible one-to-one matches, which totals

$300 = \binom{25}{2}$ combinations. The equations of RD for two strategies are only the two coupled non-linear differential equations. Combining these with the population conservation condition, it becomes a one-variable nonlinear (cubic) differential equation. One can easily find the fixed points and their stabilities [7, 12]. The RD results belong to four cases as shown in Fig. 2: (a) 189 win/lose directed links, (b) 3 unstable links, (c) 40 stable links, and (d) 68 neutral constant links. Among them, only the win/lose overwhelmed case, which is 63% of the matches, is independent of the initial conditions.

### 3. Win/lose Directed Network and Dynamic Motifs

In order to simplify and intuitively understand the competitive relations among the strategies of IPDG, we construct win/lose directed networks by using only the 189 win/lose directed links. From this directed network, we find a giant strongly-connected component (GSCC), which consists of 10 strategies. The GSCC in the directed networks naturally means the existence of directed loops (circulation). In this win/lose network, the loop is a win/lose circulation, such as rock-paper-scissors. One can find win/lose circulations among several strategies, including the smallest three in the GSCC. We analyze the smallest win/lose circulation and its stability. Extending the analysis, we enumerate all possible $2300 = \binom{25}{3}$ combinations among the three strategies and analyze their fixed points and stabilities, i.e., their phase portraits. According to network theory terminology, the simple building block among the three nodes in complex networks is called a "network motif" [13]. Because we consider four different types of links and their dynamic characteristics, it is the dynamic motif among the three strategies. These analyses of the building block, i.e., the dynamic motif, among the strategies in game theory help us to understand more intuitively the riches of dynamic behaviors and the interplay among them.

## III. RESULTS

1. **Bow-tie Structure of the Win/Lose Directed Network**

   Incorporating the 189 win/lose directed links, we construct a win/lose directed network among the 25 strategies, as shown in Fig. 3. The GSCC has 10 strategies. On the left side of the GSCC, 9 strategies are the in-components, and the 6 on the right are the out-components. Because the direction of the link, for example, heading from A to B, represents the strategy, B overwhelms A. The strategies in the out-components are the superior ones in the IPDG. In this network, the grim trigger and the TFT are on the right end, which means that they are the most dominant strategies in the IPDG for a given payoff [6, 7]. In the enlarged inset of Fig. 3, one finds the win/lose circulations where the simplest one consists of three strategies such as the rock-paper-scissors. All the smallest win/lose circulations with three strategies always contain strategy No. 5, where cooperation occurs only when the previous choice is C and the opponent's is D; otherwise, it is always defeated [7].

2. **Win/Lose Circulation and Stability Analysis**

   For a more precise analysis of these win/lose circulations, in the sense of RD, we chose two sets of three strategies. The first one was $\{5, 22, 28\}$, and the other was $\{5, 18, 17\}$ according to the direction (see Fig. 4). The results of RD showed that the first set $\{5, 22, 28\}$ had an evolutionarily stable strategy (ESS), i.e., a fixed point at (0.13, 0.39, 0.48) in configuration space [Fig. 4(a)]. All the realizations that start from the vicinity of the three corners converge on a fixed point as time evolves. On the other hand, the other set $\{5, 18, 17\}$ always diverges rotating around the point at (0.11, 0.23, 0.66), i.e., anti-ESS; in other words, the point is an unstable fixed point. As time goes by, the period of oscillation grows exponentially, and the point is expelled to the boundary.

   For the three strategies, the RD equations are just three nonlinear differential equations and are reducible to two nonlinear differential equations with the population conservation condition, $x_1 + x_2 + x_3 = 1$. If there are fixed points within the range $0 < x_i < 1$, the steady-state condition gives $\dot{x}_1 =$ $\dot{x}_2 = \dot{x}_3 = 0$ and $f_1 = f_2 = f_3 = \bar{f}$. Then, one can obtain the condition with the average payoff matrix of the **P** strategies as follows:

$$\begin{pmatrix} f_1 \\ f_2 \\ f_3 \end{pmatrix} = \begin{pmatrix} P_{11} & P_{12} & P_{13} \\ P_{21} & P_{22} & P_{23} \\ P_{31} & P_{32} & P_{33} \end{pmatrix} \begin{pmatrix} x_1^* \\ x_2^* \\ x_3^* \end{pmatrix} = \begin{pmatrix} \bar{f} \\ \bar{f} \\ \bar{f} \end{pmatrix}; \quad (1)$$

i.e., $\mathbf{P}\vec{x}^* = \bar{f}\mathbf{1}$, where **1** is the column vector of ones and $\vec{x}^*$ is the fixed point. By applying the inverse matrix of the average payoff matrix **P**, one can get the fixed point. For the cases of Fig. 4, the results of the numerical integration agree with the analytical solutions. The stability and the abstract of the phase portrait can be analyzed with the determinant and trace of the Jacobian matrix of the RD equations [12]. The examples in Fig. 4 correspond to a stable spiral [Fig. 4(a)] and an unstable spiral [Fig. 4(b)], respectively.

## 3. Network Motifs and Dynamic Motifs

Extending the concept of win/lose circulations, we enumerate all possible types of relations among the three strategies. For example, the cases in Fig. 4 have relations of A wins B, B wins C, and C wins A. Reading the relation starting from the top of the triangle counterclockwise, it is Win-Win-Win denoting the relations as "WWW." Similarly, we encode in three-character words the relation "W (L)" for win (lose), "S (U)" for stable (unstable), and "N" for the neutral constancy links in Fig. 2. The five types of relations (links) when the counterclockwise reading direction is considered, give 30 different network motifs in theory. However, for the strategies of the IPDG with a given payoff matrix, only 22 types of motifs among 30 appear [13]. All 22 existent dynamic motifs are summarized in Table 3 with their frequencies. The total combinations of choosing 3 from 25 is $2300 = \binom{25}{3}$. The most frequent motif "WWL" appears 547 times among the 2300 (23.8%), which corresponds to a situation in which there is one superior strategy. The win/lose circulation WWW quite rare happening only 4 times. Because the neutral constancy link is quite abundant (68/300 = 0.23), NNN is not very rare (74 times,

3.2%) compared to WWW.

For these 22 types of dynamic motifs, we analyze the existence of fixed points within the range of configuration space and phase portrait that describes the dynamic behaviors. On some occasions, even the motif notation is the same; the dynamics behavior is totally different as in the example in Fig. 4. All the existent dynamic motifs are displayed in Fig. 5. The directions of the flows are represented by black arrows, and the velocities are visualized with a range of colors from purple (slow) to red (fast). If there are fixed points, they are denoted by the red-filled symbol. The phase portraits of some dynamic motifs contain a stable fixed point inside the configuration space. Here, we ignore trivial cases on the edges where at least one population is zero.

## IV. SUMMARY AND CONCLUSIONS

We investigated the relations between the strategies of iterated prisoner's dilemma games (IPDG) by using a directed network to display the results of replicator dynamics. By incorporating the win/lose relation between strategies, a win/lose network is constructed adapting a directed network concept. From the analysis of the giant strongly-connected component, we find win/lose circulations like the rock-paper-scissors game and analyze the fixed point and its stability by applying a nonlinear dynamics analysis. Extending the network motif concept in directed networks, we introduce dynamic motifs, which describe the relations among the three strategies and the dynamic behaviors in the population configuration space. The five types of relations, two directed and three undirected, between two strategies generate 30 possible dynamic motifs. Among them, only 22 dynamic motifs appear in the IPDG with a given payoff matrix and a limited memory size. Visualization using directed networks and its motif analysis is a useful method to make the complex dynamic behaviors among the strategies simple so that the dynamics can be intuitively understood. This is achieved even when starting from the countless possibilities of the initial conditions. The dynamic motifs are the building blocks for understanding the dynamic behaviors of the strategies, just as network motifs are in complex networks.

ACKNOWLEDGMENTS

We would like to thank Pan-Jun Kim and Seung Ki Baek for helpful comments and discussions. This work was supported by a National Research Foundation of Korea (NRF) grant funded by the Ministry of Science, ICT & Future Planning (No. 2012R1A1A1012150).

(2002).

**Tables**

Table 1. Payoff matrix for the player (opponent) in a PD game.

|  |  | Opponent | |
|---|---|---|---|
|  |  | C | D |
| Player | C | R ( R ) | S ( T ) |
|  | D | T ( S ) | P ( P ) |

Table 2. (Color online) Summary of the 32 strategies. The duplicated strategies are marked in red.

| Strategy number | Bitwise representation | Nickname | Strategy number | Bitwise representation | Nickname |
|---|---|---|---|---|---|
| 0 | D\|DDDD | All D | 16 | D\|DDDC |  |
| 1 | C\|DDDD | All D FC | 17 | C\|DDDC |  |
| 2 | D\|CDDD | All D | 18 | D\|CDDC | Pavlov FD |
| 3 | C\|CDDD | Grim trigger | 19 | C\|CDDC | Pavlov |
| 4 | D\|DCDD | All D | 20 | D\|DCDC | Tree frog FD |
| 5 | C\|DCDD |  | 21 | C\|DCDC | Tree frog |
| 6 | D\|CCDD | All D | 22 | D\|CCDC | Prodigal son |
| 7 | C\|CCDD | All C | 23 | C\|CCDC | All C |
| 8 | D\|DDCD | Swindler FD | 24 | D\|DDCC | CD repeater |
| 9 | C\|DDCD | Swindler | 25 | C\|DDCC | DC repeater |
| 10 | D\|CDCD | Tit for tat FD | 26 | D\|CDCC | Punisher FD |
| 11 | C\|CDCD | Tit for tat | 27 | C\|CDCC | Punisher |
| 12 | D\|DCCD |  | 28 | D\|DCCC | Rebel FD |
| 13 | C\|DCCD |  | 29 | C\|DCCC | Rebel |
| 14 | D\|CCCD | Blinder | 30 | D\|CCCC | All C FD |
| 15 | C\|CCCD | All C | 31 | C\|CCCC | All C |

Table 3. All 22 of the existent dynamic motifs and their frequencies and percentages.

| rank | motif | frequency | Percentage (%) | rank | motif | frequency | Percentage (%) |
|---|---|---|---|---|---|---|---|
| 1 | WWL | 547 | 23.8 | 12 | NLS | 40 | 1.7 |
| 2 | LWN | 333 | 14.5 | 13 | WLU | 29 | 1.3 |
| 3 | WWN | 253 | 11.0 | 14 | SSS | 17 | 0.7 |
| 4 | WLN | 252 | 11.0 | 15 | NNS | 17 | 0.7 |
| 5 | LWS | 231 | 10.0 | 16 | NNU | 13 | 0.6 |
| 6 | NWS | 177 | 7.7 | 17 | WWU | 12 | 0.5 |
| 7 | SSN | 95 | 4.1 | 18 | LWU | 7 | 0.3 |
| 8 | SSW | 84 | 3.7 | 19 | WWW | 4 | 0.2 |
| 9 | NNN | 74 | 3.2 | 20 | NWU | 3 | 0.1 |
| 10 | NNW | 63 | 2.7 | 21 | UUN | 2 | 0.1 |
| 11 | WWS | 46 | 2.0 | 22 | NLU | 1 | 0.0 |

**Figures & Figure Captions**

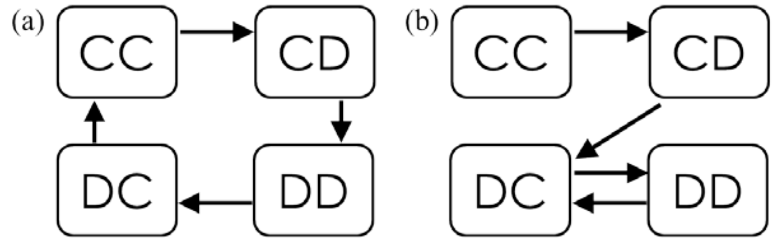

Fig. 1. State transition graph between two strategies. (a) TFT versus "Tree frog" has no transient state and there are four periods. (b) GT versus "CD repeater" has transient states before entering the period-two recurrent state.

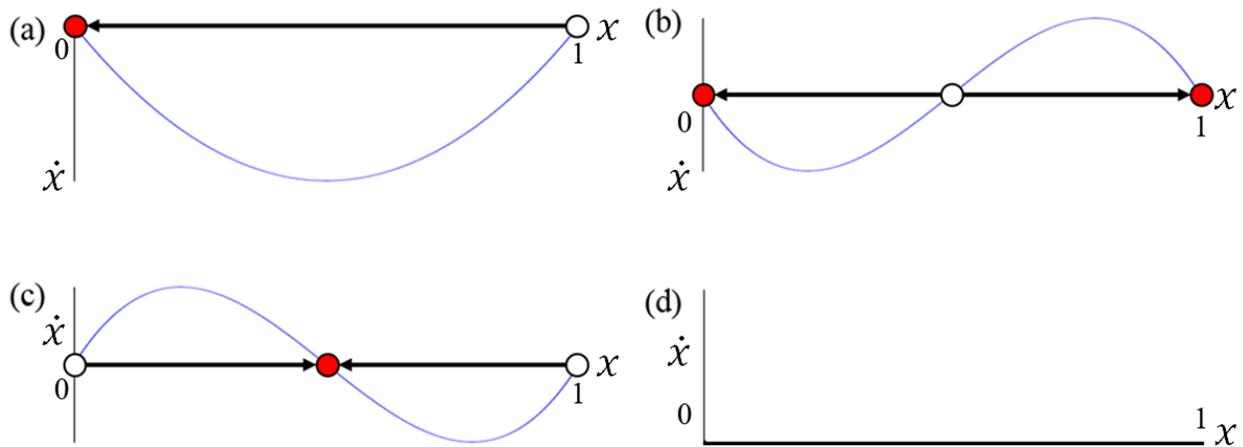

Fig. 2. (Color online) Four possible cases of replicator dynamics for one-to-one matches: (a) win/lose, (b) unstable, (c) stable, and (d) constant cases. The stable and the unstable fixed points are marked in red-filled and white-unfilled symbols, respectively.

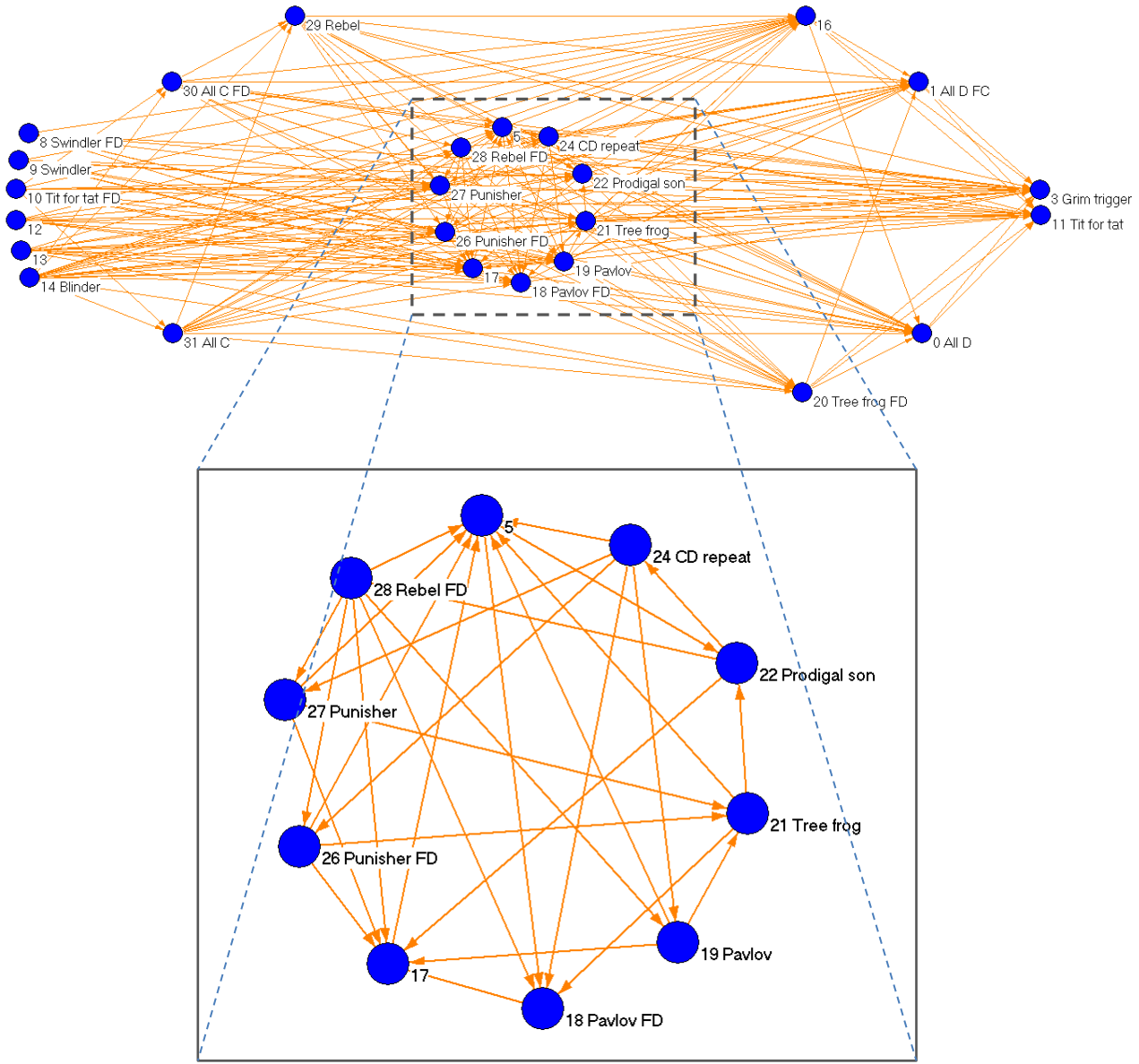

Fig. 3. (Color online) Bow-tie network structure and the giant strongly-connected component. The bow-tie structure shows that the strategies on the left side are the out-components with "lose" connections, and the strategies on the right side are the in-components with "win" connections. There are 10 strategies in the giant strongly-connected component (GSCC). In the GSCC, we find a win-lose circulation among the strategies.

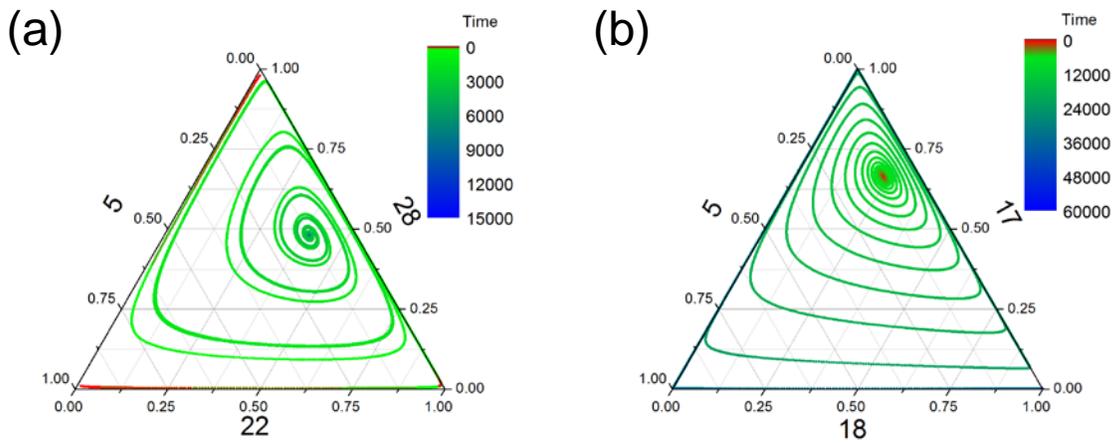

Fig. 4. (Color online) Evolution dynamics in the win-lose circulation: (a) stable (ESS) and (b) unstable fixed point (anti-ESS). Red indicates the early step, and blue is the later one. For (a), three different initial states, (0.98, 0.01, 0.01), (0.01, 0.98, 0.01), and (0.01, 0.01, 0.98), are used. For (b), the simulation starts from the vicinity of the unstable fixed point.

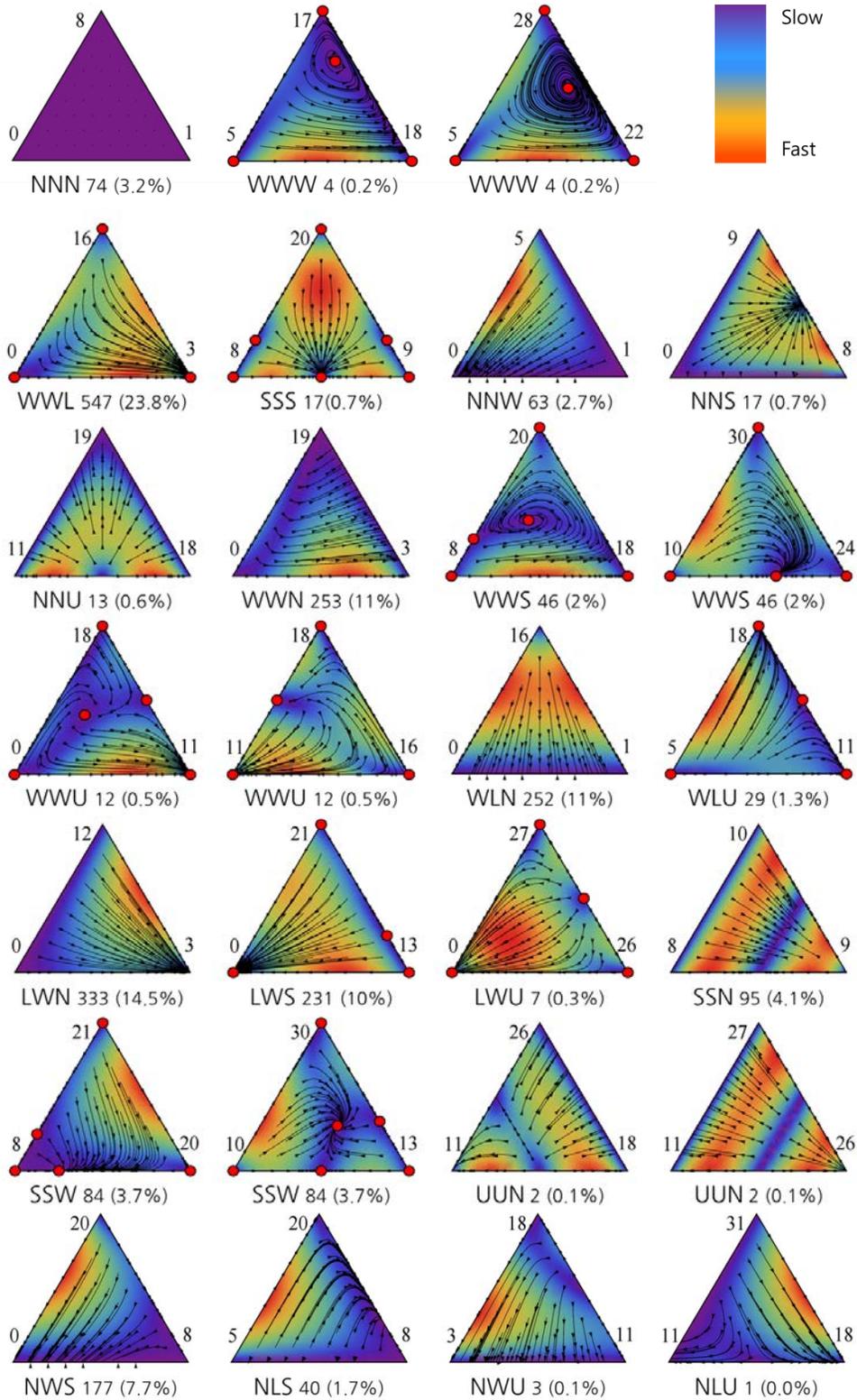

Fig. 5. (Color online) All possible dynamic results of the motifs among the three strategies. The stable fixed points are marked by red-filled circles, and the velocities are represented as shown in the top right; reddish is fast, bluish is slow, and purple indicates almost stopped.